%% file: main.tex
\documentclass[structabstract]{aa}
\usepackage{txfonts}
\usepackage{graphicx}
\usepackage{natbib}
\bibpunct{(}{)}{;}{a}{}{,} 

\usepackage{color}
\usepackage{longtable}
\usepackage{lscape}

\begin{document}

\title{A survey of T Tauri stars with AKARI toward the Taurus-Auriga region}

\subtitle{}

\author{Satoshi Takita \inst{1,2}
  \and Hirokazu Kataza \inst{2}
  \and Yoshimi Kitamura \inst{2}
  \and Daisuke Ishihara \inst{3}
  \and Yoshifusa Ita \inst{4}
\thanks{Present address is Astronomical Institute, Graduate School of Science, Tohoku University, 6-3 Aramaki, Aoba-ku, Sendai, 980-8578, Japan}
  \and Shinki Oyabu \inst{2}
\thanks{Present address is Subaru Telescope, National Astronomical Observatory of Japan, 650 North A'ohoku Place, Hilo, HI 96720, USA}
  \and Munetaka Ueno \inst{2}}


\institute{
  Department of Space and Astronautical Science, The Graduate University for Advanced Studies (Sokendai), 3-1-1 Yoshinodai, Chuo, Sagamihara, Kanagawa 229-8510, Japan
  \and 
  Institute of Space and Astronautical Science, Japan Aerospace Exploration Agency, 3-1-1 Yoshinodai, Chuo, Sagamihara, Kanagawa 229-8510, Japan
  \and 
  Division of Particle and Astrophysical Sciences, Nagoya University, Furo-cho, Chikusa-ku, Nagoya, 464-8602, Japan
  \and 
  National Astronomical Observatory of Japan, 2-21-1 Osawa, Mitaka, Tokyo 181-8588, Japan
  }

\date{Received 2 November 1992 / Accepted 7 January 1993}

\abstract
{
AKARI, the first Japanese infrared astronomical satellite, has completed an All-Sky Survey at the mid- to far-infrared wavelengths with greater spatial resolutions and sensitivities than those of the previous survey with Infrared Astronomical Satellite (IRAS).
}
{
We search new T Tauri star (TTS) candidates with the mid-infrared (MIR) part of the AKARI All-Sky Survey at 9 and 18 $\mu$m wavelengths.
}
{
We used the point source catalogue (PSC), obtained by the Infrared Camera (IRC) on board AKARI.
We combined the 2MASS PSC and the 3rd version of the USNO CCD Astrograph Catalogue (UCAC) with the AKARI IRC-PSC, and surveyed 517 known TTSs over a 1800-square-degree part of the Taurus-Auriga region to find criteria to extract TTSs.
We considered asymptotic giant branch (AGB) stars, post-AGB stars, Planetary Nebulae (PNe), and galaxies, which have similar MIR colours, to separate TTSs from these sources.
We finally searched new TTS candidates from AKARI IRC-PSC in the same Taurus-Auriga region.
}
{
Of the 517 known TTSs, we detected 133 sources with AKARI: 46 sources were not detected by IRAS.
Based on the colour-colour and colour-magnitude diagrams made from the AKARI, 2MASS, and UCAC surveys, we propose the criteria to extract TTS candidates from the AKARI All-Sky data, and 68/133 AKARI detected TTSs have passed these criteria.
On the basis of our criteria, we selected 176/14725 AKARI sources as TTS candidates which are located around the Taurus-Auriga region.
Comparing these sources with SIMBAD, there are 148 previously identified sources including 115 Young Stellar Objects (YSOs), and 28 unidentified sources.
}
{
Based on SIMBAD identifications, we take the TTS-identification probability using our criteria to be $\sim$75 \%.
We find 28 TTS candidates, of which we expect $\sim$21 to be confirmed once follow-up observations can be obtained.
Although the probability of $\sim$75 \% is not so high, it is affected by the completeness of the SIMBAD database, and we can search for TTSs over the whole sky, over all star forming regions.
}


\titlerunning{Taurus TTSs with AKARI}
\authorrunning{Takita et al.}

\maketitle

\input{chapter/introduction}
\input{chapter/akari}
\input{chapter/target}
\input{chapter/colour}

\input{chapter/new_tts}

\input{chapter/acknowledgment}

\input{chapter/reference}
\input{chapter/table1}
\input{chapter/table2}

\end{document}

%% file: chapter/introduction.tex
\section{INTRODUCTION}

T Tauri stars (TTSs) are low-mass pre-main sequence (PMS) stars with ages of $\sim 10^6$--$10^7$ yrs \citep{als1988}.
TTSs were traditionally identified with their strong H$\alpha$ emission line toward their natal molecular clouds, i.e., classical TTSs (CTTSs) with ages of $\sim 10^6$ yr.
On the other hand, X-ray observations, such as the Einstein Observatory and ROSAT surveys, have discovered another class of PMS stars (e.g., \citealt{neuhauser1995a}).
Since these sources have the ``weak'' H$\alpha$ emission line with equivalent widths of $\le$ 10 \AA, they are called weak-line TTSs (WTTSs) and have ages of $\sim 10^7$ yr.
The youth of WTTSs was confirmed by strong Li \textsc{i} absorption lines, because lithium is easily destroyed in the stellar atmosphere with high temperature.
In contrast to the CTTS case, many WTTSs are found outside the molecular clouds.
This situation is interpreted as follows:
The natal molecular clouds have been already dispersed at $\sim 10^7$ yr, or the stars have left the natal clouds owing to their motions with a few mas/yr.

The circumstellar disks around TTSs are believed to be the birthplaces of planets.
The pioneering studies of the disks are the IRAS and millimetre continuum surveys towards the Taurus-Auriga region \citep{strom1989,beckwith1990}.
\citet{strom1989} found that about a half of CTTSs have excess emission at the infrared (IR) wavelength stronger than that expected from their photospheres.
The excess emission is well interpreted as the thermal emission from the circumstellar disks heated by the central stars and/or mass accretion.
On the other hand, most WTTSs lack such excess emission.
Even the recent IR observations with the \textit{Spitzer} Space Telescope have confirmed such a low detection rate of the excess emission towards WTTSs.
\citet{silverstone2006} showed that 5 of 74 young stars (3--30 Myr) have strong IR excess, but they have spectral energy distributions consistent with CTTSs.
\citet{padgett2006} observed 83 WTTSs outside their natal clouds, and found only 5 WTTSs have excess emission.
The largest \textit{Spitzer} survey of WTTSs so far was done by \citet{cieza2007}, which observed more than 230 WTTSs located in the Ophiuchus, Lupus, and Perseus molecular clouds.
Their data indicate that $\sim$20 \% of the WTTSs have IR excess emission, but no IR excess for the stars older than 10 Myr.
Consequently, it is most likely that the disk dissipation time scale is about 10 Myr.
However, IRAS did not have enough sensitivities to detect the excess emission of WTTSs, and \textit{Spitzer} cannot cover ``all'' WTTSs.
Therefore, we need unbiased and high-sensitivity surveys to study the WTTS disks with good statistics.

%% file: chapter/akari.tex
\section{AKARI all-sky data}

\subsection{AKARI IRC All-Sky Survey}

AKARI is the first Japanese infrared astronomical satellite dedicated to infrared astronomy \citep{murakami2007}.
One of the major observation programs of AKARI is an All-Sky Survey at the mid- to far-infrared wavelengths with 6 photometric bands.
AKARI has a higher sensitivity, a higher spatial resolution, and a wider wavelength coverage than those of the previous IRAS survey.
The mid-infrared (MIR) survey has been carried out with the \textit{S9W} (9 $\mu$m) and \textit{L18W} (18 $\mu$m) bands using the Infrared Camera (IRC; \citealt{onaka2007}).
The 5 $\sigma$ detection limit for a point source is estimated to be 50 and 120 mJy at the \textit{S9W} and \textit{L18W} bands, respectively.
The spatial resolution is around 5$''$.
More than 96 \% of the entire sky has been observed with the two bands.
The first version of the AKARI IRC point source catalogue (hereafter IRC-PSC) was publicly released in March 2010.
The analysis of this paper is based on the 1st version of the IRC-PSC.
The details of the AKARI IRC All-Sky Survey and its data reduction processes are described in \citet{ishihara2010}.

\subsection{Comparison with the NIR 2MASS and optical UCAC catalogues}

We compared the IRC-PSC with the 2MASS PSC by a near-infrared (NIR) survey \citep{2mass} using a simple positional correlation method.
We used a positional tolerance of 5$''$, which is the spatial resolution of the AKARI IRC All-Sky Survey.
More than 99 \% of the AKARI sources agree well with those in the 2MASS PSC within the accuracy.
We also compared the IRC-PSC with the 3rd version of the USNO CCD Astrograph Catalogue (UCAC) by an optical survey \citep{ucac3} in the same way.
About 70 \% of the AKARI sources have optical counterparts within the 5$''$ accuracy.
This relatively low cross-identification rate comes from the limited magnitude range in the UCAC survey:
Most nearby ($<$ 100 pc) stars are saturated, and distant ($>$ 1 kpc) or heavily reddened stars have no UCAC entry because of its sensitivity.
Since the UCAC catalogue contains ``stars'' with the magnitude range of $R$ = 7.5--16.3 in a 579--642 nm band, it is useful to search stars at distances of about 100 pc, which is the typical distance to nearby molecular clouds
(for example, if we put the Sun at 140 pc, the magnitude becomes $\sim$11).

%% file: chapter/target.tex
\section{AKARI IRC observations of the previously known T Tauri stars in the Taurus-Auriga region}

\subsection{Previously known members in the Taurus-Auriga region}
\begin{figure*}[ht]
\begin{center}
\includegraphics{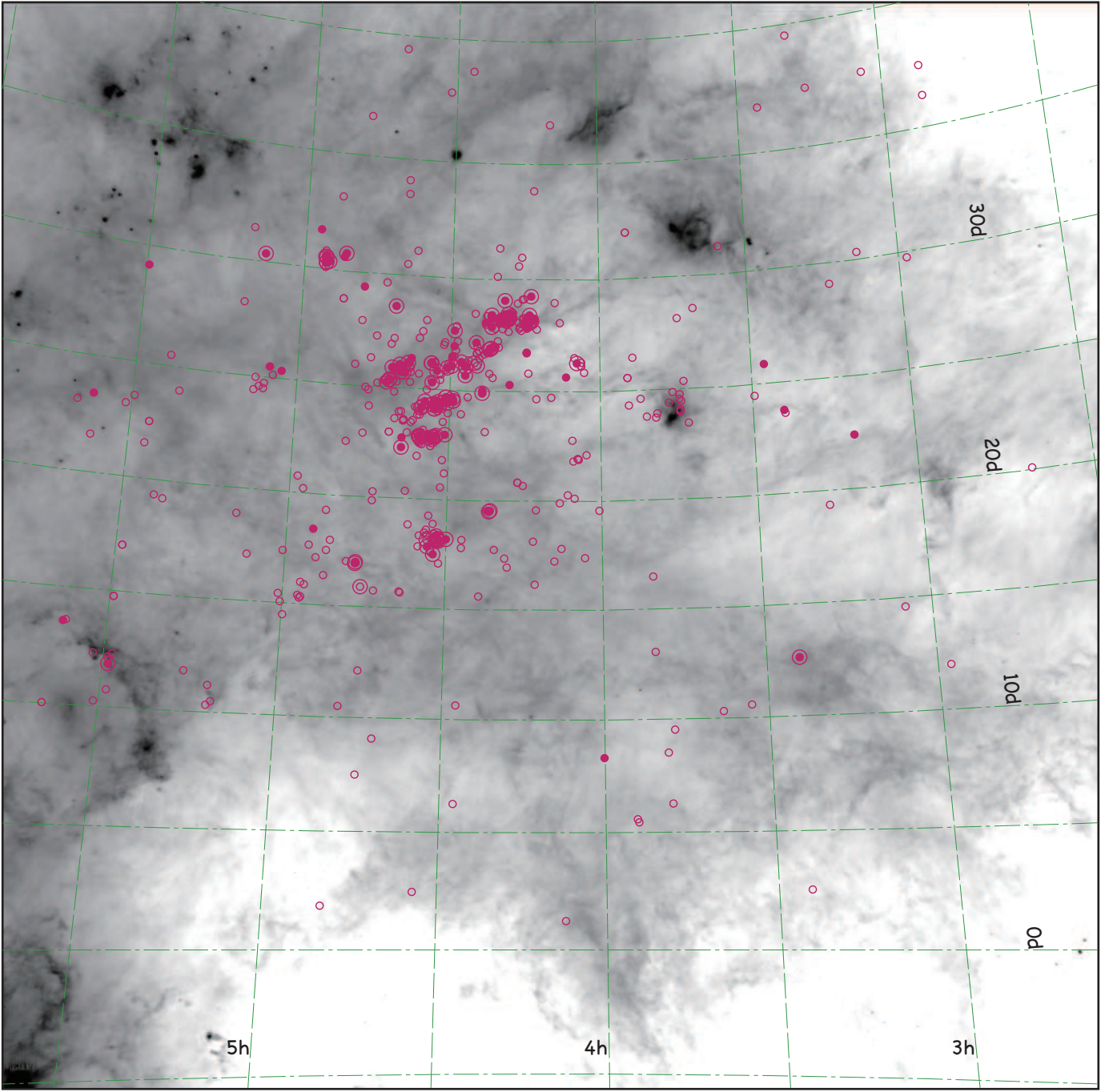}
\caption{Distribution of the target TTSs (small open circles).
The small filled circles indicate the TTSs detected with AKARI and the large open ones mean the IRAS detection, on the IRAS 100 $\mu$m map.
}
\label{fg:ttspos}
\end{center}
\end{figure*}

We selected a 1800-square-degree ($2^{\rm h}40^{\rm m} < {\rm R.A.} < 5^{\rm h}40^{\rm m}$ and $0^\circ < {\rm Dec} < 40^\circ$) part of the Taurus-Auriga region for this study.
The Taurus-Auriga region is a well-studied low-mass star-forming one at a close distance of $\sim$ 140 pc with hundreds of pre-main-sequence (PMS) stars.
We first compiled known Taurus PMS members from \citet{rebull2010}, \citet{kenyon2008}, \citet{guedel2007}, \citet{beckwith1990}, and \citet{strom1989}, which are associated with the Taurus molecular cloud.
Second, we picked up TTSs from the PMS objects as follows:
(1) Class II/III objects including the `new', `probable', and `possible' members in \citet{rebull2010};
(2) Objects listed in \citet{kenyon2008} that were confirmed as Class II/III objects on the basis of the SIMBAD database and their references;
(3) Objects which are labelled as CTTSs or WTTSs in \citet{guedel2007}.
Note that \citet{beckwith1990} and \citet{strom1989} listed only TTSs.
Furthermore, we added TTSs which are discovered by the Roentgen satellite (ROSAT) and distributed from the cloud \citep{li1998, magazzu1997, wichmann1996}.
In our TTS list, we considered close ($< 5''$; the spatial resolution of the AKARI IRC All-Sky Survey) binaries as a single source.
Finally, we got a catalogue of 517 TTSs in the selected region.
The distribution of our input TTSs is shown in Figure \ref{fg:ttspos}.

\subsection{Statistical Properties of the previously known T Tauri stars detected with AKARI}

\begin{figure}[ht]
\begin{center}
\includegraphics{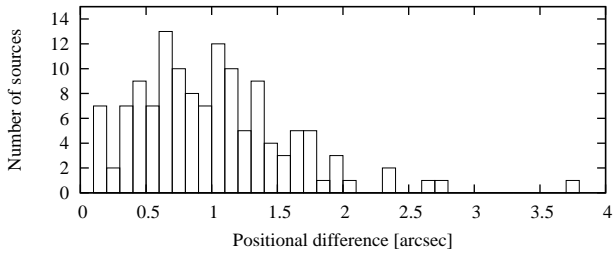}
\caption{
Histogram of the positional differences between the TTSs detected with AKARI in the IRC-PSC and those in the previous catalogues.
The size of the positional difference bin is 0.1 arcsec.
}
\label{fg:distance}
\end{center}
\end{figure}

We totally detected 133 TTSs with AKARI within the positional accuracy of 5$''$, i.e., the resolution of the AKARI IRC All-Sky Survey: 127 and 95 sources were detected at the \textit{S9W} and \textit{L18W} bands, respectively.
There are 6 sources which were detected only with the \textit{L18W} band:
  3 sources (IRAS 04200+2759, IRAS 04295+2251, and UX Tau) were observed/detected only once at the \textit{S9W} band, so the IRC-PSC does not contain these sources, and the other 3 sources were not detected at all.
The positions of almost all the AKARI detected sources agree well with those in the previous catalogues within the uncertainty of 3$''$ (see Figure \ref{fg:distance}).
Although GI Tau has larger separations than 3$''$, it still satisfy the positional accuracy of AKARI IRC-PSC.
The photometric data of the AKARI detected TTSs are listed in Table \ref{tbl:ttsdata}.

\begin{figure}[ht]
\begin{center}
\includegraphics{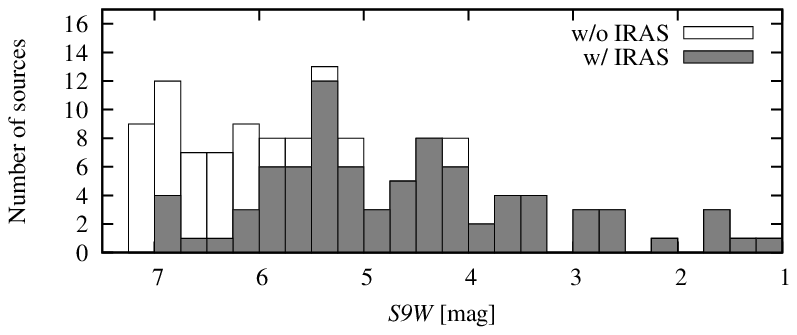}\\
\includegraphics{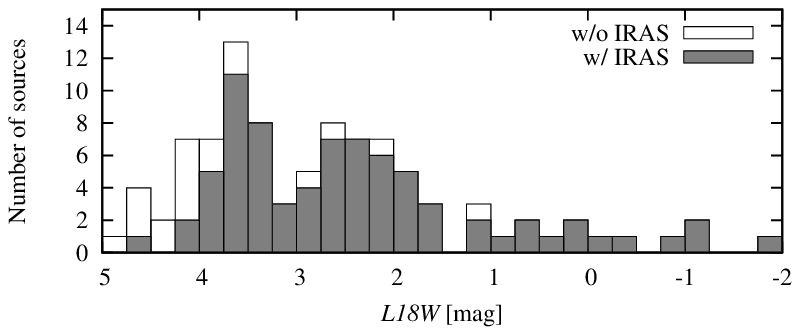}
\caption{
Histogram of the \textit{S9W} (\textit{top}) and \textit{L18W} (\textit{bottom}) magnitudes of the TTSs detected with AKARI.
The filled and open bars indicate the sources that were detected and not detected with IRAS, respectively.
The size of the magnitude bin is 0.25 magnitude.
}
\label{fg:magnitude}
\end{center}
\end{figure}

\begin{figure}[ht]
\begin{center}
\includegraphics{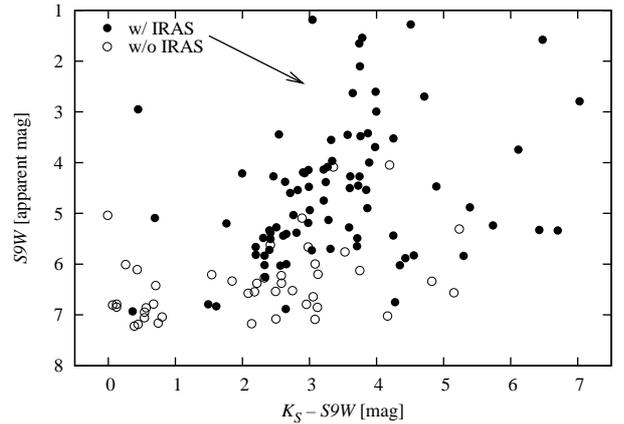}
\caption{
($K_S - S9W$) v.s. ($S9W$) colour-magnitude diagram of the AKARI detected PMS stars.
The filled and open circles indicate the sources that were detected and not detected with IRAS, respectively.
The arrow shows the interstellar extinction vector of $A_V$ = 20 mag, using the \citet{weingartner2001} Milky Way model of $R_V$ = 3.1.
}
\label{fg:cmd0}
\end{center}
\end{figure}

In the 133 TTSs, 46 faint sources have no counterparts in the IRAS PSC (FQUAL12 = 3 or FQUAL25 = 3) with a searching radius of 60$''$, the IRAS resolution, as shown in Figure \ref{fg:magnitude}.
These `new detections' are due to the higher sensitivity and spatial resolution of AKARI than those of IRAS.
Actually, Figure \ref{fg:magnitude} shows that the detection limit for the previously known Taurus TTSs at the \textit{S9W} band has been improved:
IRAS could detect almost all the TTSs that have brighter \textit{S9W} and \textit{L18W} band magnitudes than 6 and 4, respectively, but only about a quarter of the fainter sources were detected with IRAS.
Figure \ref{fg:cmd0} shows the ($K_S - S9W$) v.s. ($S9W$) colour-magnitude diagram of the AKARI detected sources.
Since about the half of the sources that were not detected with IRAS have small colours with $K_S - S9W < 1$ and most of the IRAS detected TTSs have larger colours with $K_S - S9W > 2$, there seems to exist a gap at $K_S - S9W \sim 0.5$.
The sources with $K_S - S9W < 1$ are most likely to be WTTSs, because they have weak H$\alpha$ emission (except HT Tau, whose H$\alpha$ equivalent width is not given) and are located near the periphery of the clouds or outside the clouds.
AKARI has succeeded in detecting WTTSs with weak excess emission at the sensitive \textit{S9W} band.

There are 18 TTSs which are not catalogued in the IRC-PSC within a 5$''$ searching radius, but are catalogued in the IRAS PSC within 60$''$ radius.
Of these sources, 15 sources have AKARI counterparts within the positional accuracy of the IRAS, but the remaining 3 sources have no counterparts in the IRC-PSC.
IRAS 04302+2247 was observed once or four times at the \textit{S9W} or \textit{L18W} bands, respectively, but not detected.
Although this source is a well-known TTS with an edge-on disk, it is faint at the \textit{S9W} and \textit{L18W} bands, suggesting the presence of an inner gap in the disk..
Indeed, it has bright 24 $\mu$m magnitude of 3.57 but faint 8 $\mu$m one of 9.71 \citep{rebull2010}.
IRAS 04216+2603 was observed four times and detected only once at the \textit{L18W} band, and not observed at the \textit{S9W} band at all.
Since the IRC-PSC requires more than one detection for a `real' point source to reject a moving object or a fake one, the source is not catalogued in the IRC-PSC.
Finally, DM Tau was detected twice at the \textit{L18W} band, but the positions in the two independent images have a larger difference than 5$''$ between each other under the current positional accuracy, i.e., no entry in the IRC-PSC.

\begin{figure}[ht]
\begin{center}
\includegraphics{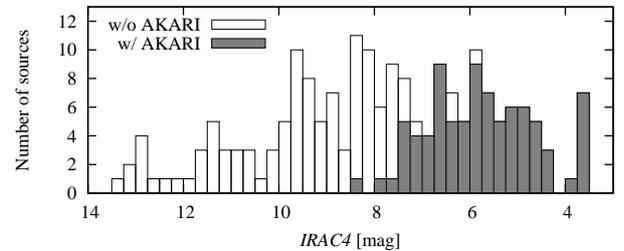}
\caption{
Histogram of the \textit{IRAC4} magnitudes of the PMS stars which are listed in Table 4 by \citet{rebull2010}.
The filled and open bars indicate the sources that were detected and not detected with AKARI, respectively.
The size of the magnitude bin is 0.25 magnitude.
}
\label{fg:sst}
\end{center}
\end{figure}

Although Spitzer didn't cover the entire region we searched, sources inside the Taurus molecular cloud and some other sources outside the cloud have been also observed with Spitzer (e.g., \citealt{rebull2010}).
Since Spitzer has higher sensitivity than AKARI All-Sky Survey, Spitzer has discovered more faint sources as shown in Figure \ref{fg:sst}, which shows the histogram of the IRAC4 (8 $\mu$m) magnitudes of the detected TTSs with Spitzer.
The AKARI All-Sky Survey could detect more than $\sim$90\% of the sources with magnitude brighter than 7.5 at the IRAC4 band, which agrees well with the detection limit of the AKARI \textit{S9W} band.

%% file: chapter/colour.tex
\section{How to extract TTS candidates from the AKARI All-Sky data}

\subsection{Other types of sources in the whole sky}
Since the following types of sources are known to have similar colours to those of the TTSs, we should reveal the colour properties of the sources.
We considered the four additional catalogues of
(1) asymptotic giant branch (AGB) stars of 126 carbon and 563 OH/IR stars \citep{bertre2003},
(2) 326 Post-AGB stars \citep{szczerba2007},
(3) 1143 Planetary Nebulae (PNe) \citep{acker1994}, and
(4) 2907 extragalactic objects which have brighter flux density than 100 mJy at the IRAS 12 $\mu$m band (the NASA Extragalactic Database).
In contrast to the TTS case, we examined these sources in the whole sky because of a small number of the sources toward the Taurus-Auriga region.

\subsection{Colour-colour diagrams}

\begin{figure}[ht]
\begin{center}
\includegraphics{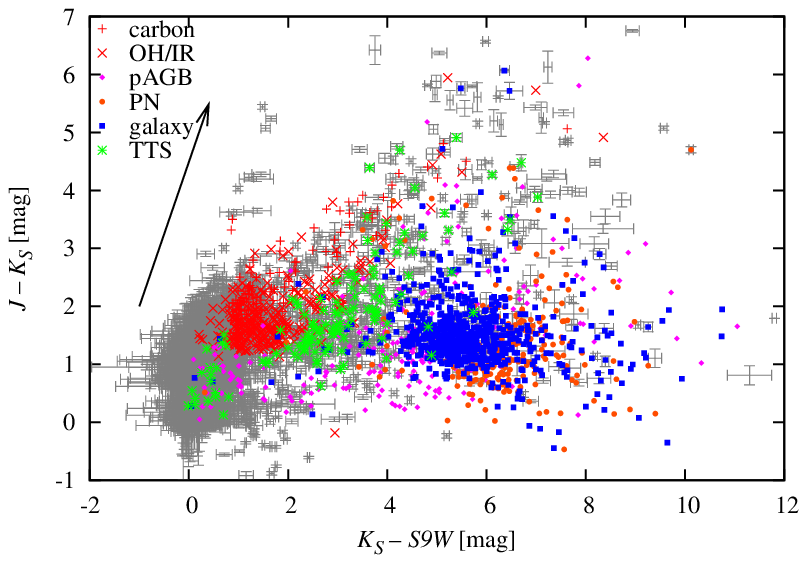}\\
\includegraphics{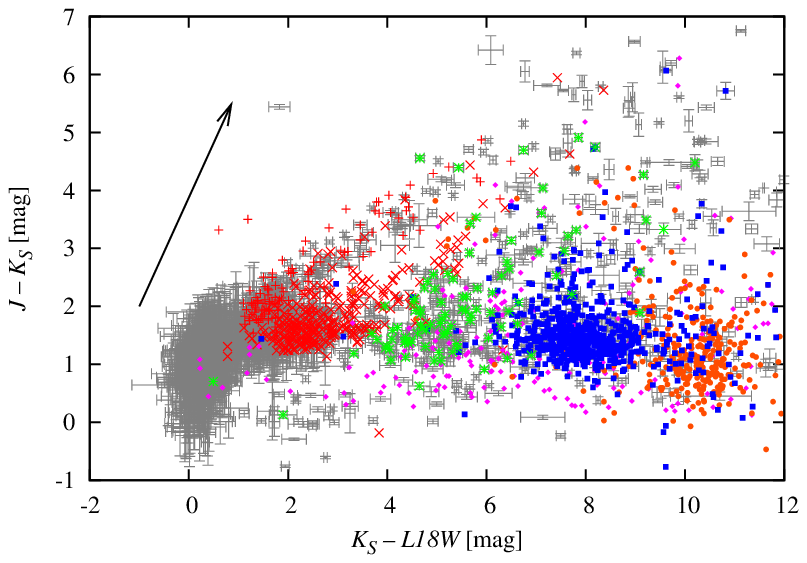}
\caption{\textit{Top}: ($K_S - S9W$) v.s. ($J - K_S$) colour-colour diagram.
The grey dots with error bars indicate all the AKARI point sources in the selected Taurus-Auriga region.
The red plus and cross symbols show carbon and OH/IR stars in the whole sky, respectively.
The magenta diamonds, the orange circles, and the blue squares are post-AGB stars, PNe, and extragalactic objects in the whole sky, respectively.
The green stars mean the TTSs in the Taurus-Auriga region.
The arrow shows the interstellar extinction vector of $A_V$ = 20 mag, estimated from the \citet{weingartner2001} Milky Way model of $R_V$ = 3.1.
\textit{Bottom}: ($K_S - L18W$) v.s. ($J - K_S$) colour-colour diagram.
The symbols and arrow are the same as in the \textit{top} panel.
}
\label{fg:ccd1}
\end{center}
\end{figure}

Figure \ref{fg:ccd1} shows the ($K_S - S9W$) v.s. ($J - K_S$) and ($K_S - L18W$) v.s. ($J - K_S$) colour-colour diagrams, where the \textit{S9W} and \textit{L18W} excess emission can be clearly recognized.
One object, 1RXS J032409.7$+$123745, does not seem to have excess emission at both the \textit{S9W} and \textit{L18W} bands among the TTSs in the Taurus-Auriga region.
Although this source is listed as a WTTS with spectral type of K2 \citep{li1998}, it seems to be a Li-rich giant star because of high luminosity ($J \sim$ 4 and $V \sim$ 6) not expected for a K2 dwarf star at a distance of $\sim$ 140 pc.
On the other hand, the other types of the sources, which are stars surrounded with dust, also have significant IR excess emission.
Therefore, we need to separate these sources from TTSs in the IRC-PSC.

\begin{figure}[ht]
\begin{center}
\includegraphics{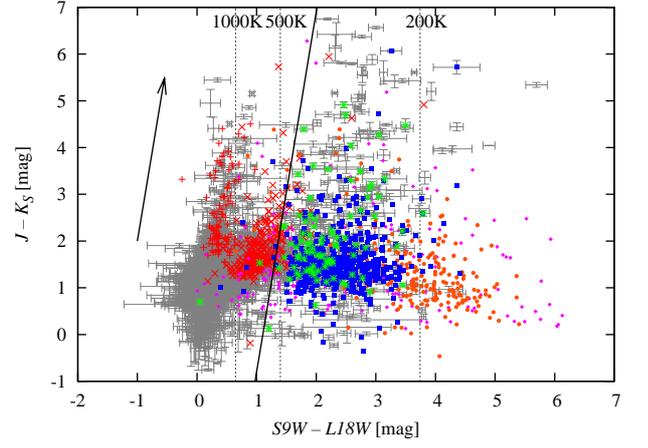}
\caption{($S9W - L18W$) v.s. ($J - K_S$) colour-colour diagram.
The symbols and arrow are the same as in Figure \ref{fg:ccd1}.
The black solid line indicates our criterion (1).
The black broken lines indicate $S9W - L18W$ colours of 200, 500, and 1000 K blackbody.
}
\label{fg:ccd3}
\end{center}
\end{figure}

First, we separate almost all carbon and OH/IR stars from the TTSs in the ($S9W - L18W$) v.s. ($J - K_S$) colour-colour diagram shown in Figure \ref{fg:ccd3}.
Since an AGB star typically has hot dust near its photosphere, the ($S9W - L18W$) colour, which represents the dust temperature, becomes \textit{blue}.
On the other hand, since the majority of the dust around a TTS is cold, the MIR colour is \textit{red}.
To remove about 80 and 70 \% of carbon and OH/IR stars, we propose the first criterion as
\begin{equation}
J - K_S \le 7.7 \times (S9W - L18W) - 8.5.
\label{eq:criterion1}
\end{equation}
However, post-AGB stars and PNe have the \textit{red} MIR colour because they have cold dust.
Furthermore, galaxies also have cold dust.
Therefore, we can separate only 15 \% of post-AGB stars, and few PNe and galaxies from TTSs.
Note that this criterion is valid for the sources that were detected at both the \textit{S9W} and \textit{L18W} bands.
We do not remove the sources which were detected at only one band.

\subsection{Colour-magnitude diagram}

\begin{figure}[ht]
\begin{center}
\includegraphics{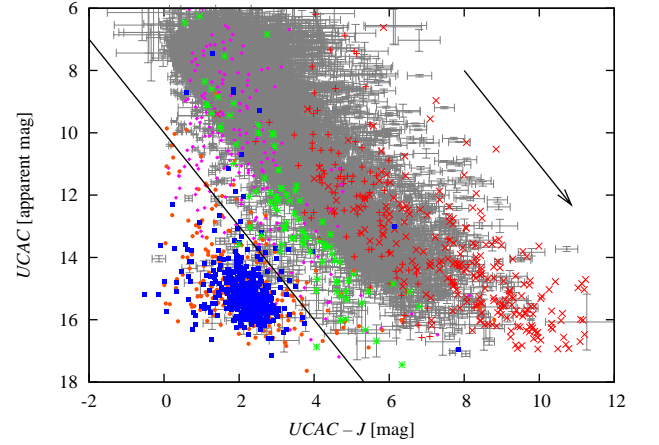}
\caption{($UCAC - J$) v.s. ($UCAC$) colour-magnitude diagram.
The symbols are the same as in Figure \ref{fg:ccd1}.
The arrow shows the interstellar extinction vector of $A_V$ = 5 mag, estimated from the \citet{weingartner2001} Milky Way model of $R_V$ = 3.1.
The black line indicates our criterion (2).
}
\label{fg:cmd1}
\end{center}
\end{figure}

Second, we can separate post-AGB stars, PNe, and galaxies from TTSs in the ($UCAC - J$) v.s. ($UCAC$) colour-magnitude diagram.
Owing to the low luminosities of PNe (white dwarfs) and galaxies at the visible wavelengths, we can remove about 90 and 97 \% of PNe and galaxies, respectively, by the following 2nd criterion:
\begin{equation}
UCAC \le 1.5 \times (UCAC - J) + 10.
\label{eq:criterion2}
\end{equation}
From this criterion, we can also get rid of 26 \% of post-AGB stars.
Furthermore, 4 and 12 \% of carbon and OH/IR stars, respectively, can be also separated.
Considering the detection limits of the IRC-PSC and the UCAC, this method seems effective only for extracting nearby ($\lesssim$ 100 pc) young sources.
Note that we never pick up TTSs with their edge-on disks, because they are not optically visible.

\begin{figure}[ht]
\begin{center}
\includegraphics{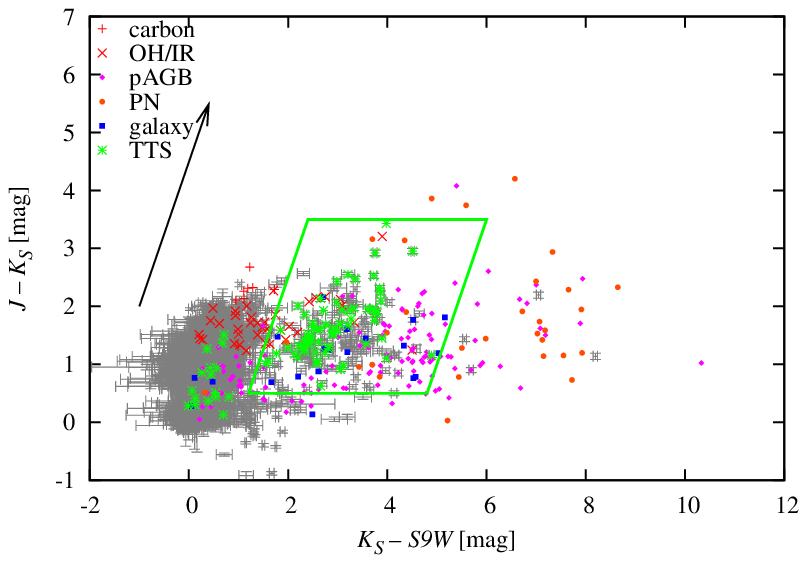}\\
\includegraphics{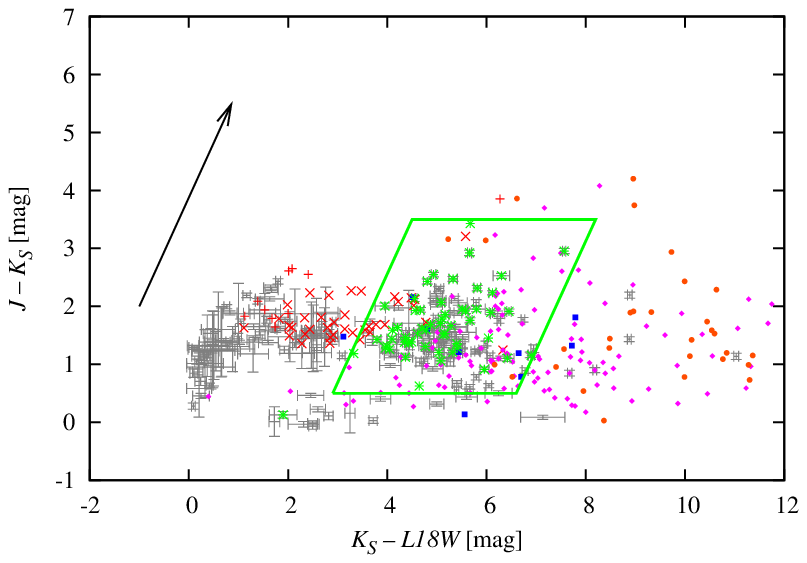}
\caption{
Colour-Colour diagrams, the same as in Figure \ref{fg:ccd1}, but for the selected sources with the criteria (\ref{eq:criterion1}) and (\ref{eq:criterion2}) in the text.
The green parallelograms indicates our criteria (3) and (4).
}
\label{fg:ccd4}
\end{center}
\end{figure}

Figure \ref{fg:ccd4} shows the same colour-colour diagrams as in Figure \ref{fg:ccd1}, but for the remaining sources that could not be removed by the criteria (\ref{eq:criterion1}) and (\ref{eq:criterion2});
most of the remaining sources with the excess emission are the TTSs.
Finally, we propose the following two criteria to pick up most of the TTSs with less contamination of other types of sources:
\begin{eqnarray}
& -11.5 \le (J-K_S) - 2.5\times(K_S-S9W) \le -2.5 \\ \nonumber
& \&\&\, 0.5 \le J - K_S \le 3.5, \\
& -12 \le (J-K_S) - 1.9\times(K_S-S9W) \le -5 \\ \nonumber
& \&\&\, 0.5 \le J - K_S \le 3.5.
\label{eq:criterion3}
\end{eqnarray}
Note that we cannot pick up TTSs which have weak MIR excess emission, because they are contaminated by field stars.
Consequently, most of AGB stars, PNe, and galaxies have been removed with our criteria.
However, 34 \% of post-AGB stars still remain, and we need follow-up observations to distinguish young stars from post-AGB stars.
Of the 133 AKARI detected TTSs, 68 sources have passed these criteria.

%% file: chapter/new_tts.tex
\section{Evaluation of the criteria}
\subsection{New TTS candidates from our criteria}
On the basis of our criteria, we selected 176 sources out of 14725 AKARI sources, which are distributed in the area between $2^{\rm h}40^{\rm m}$ and $5^{\rm h}40^{\rm m}$ in Right Ascension and between $0^\circ$ and $40^\circ$ in Declination.
We farther searched these sources in SIMBAD with a 10$''$ searching radius.
In the 176 selected sources, 21 sources could not be found in the SIMBAD database:
there were 115 YSOs, 4 Be stars, 10 AGB stars, 1 Galaxy, 18 other types of objects (mostly variable stars), and the remaining sources consist of 7 unknown objects.
Since $\sim$75 \% of the SIMBAD identified sources were YSOs, we take the TTS-identification probability based on our criteria to be 75 \%, though that is of course affected by the completeness of the SIMBAD database.

We now have found the 21 new TTS candidates in addition to the 7 unknown sources.
The infrared properties of the 28 TTS candidates are listed in Table \ref{tbl:newdata}.
Considering the TTS-identification probability of 75 \%, we expect 21 new TTSs.
We need follow-up observations to distinguish whether these sources are really TTSs or not.

\subsection{Comparison with other methods}
Many astronomers have discussed about the criteria to extract YSO candidates from their surveyed data.
Recently, \citet{evans2009} and \citet{rebull2010} surveyed YSOs in nearby clouds using Spitzer.
\citet{evans2009} surveyed 5 clouds (15.5 deg$^2$ region in total) and listed $\sim$1000 YSO candidates based on the colour-magnitude and colour-colour diagrams by 2MASS and Spitzer.
They showed that there could be 51 galaxies in their YSO candidates.
However, since their surveyed areas are highly embedded ($A_V > 2$) and are located at high galactic latitudes, they paid less attention to AGB stars.
\citet{rebull2010} surveyed a $\sim$44 deg$^2$ region of the Taurus molecular cloud, and listed 148 new candidate Taurus members based on colours and magnitudes by 2MASS and Spitzer together with the images of SDSS and CHFT.
Of these 148 new candidates, they did follow-up spectroscopic observations for about half of the sources, and confirmed 47 new Taurus candidates, 7 extragalactic objects, and 1 Be star; the remaining 93 sources await further follow-up observations.

These two studies have the higher (973/1024, \citealt{evans2009}; 47/55, \citealt{rebull2010}) YSO-identification probabilities than our probability of $\sim$75 \%.
This is mainly because we have data at only 2 bands; they usually have more than 4 bands.
Furthermore, because the detection limits of the AKARI All-Sky Survey are relatively shallower than those of the pointed observations of Spitzer, it is difficult to detect MIR faint objects.
However, since AKARI observed almost the whole sky, we can find TTS candidates toward all nearby star forming region.

%% file: chapter/acknowledgment.tex
\begin{acknowledgements}

This work is based on observations with AKARI, a JAXA project with the participants of ESA.
We gratefully acknowledge all the members of the AKARI project for their support on this project.
This publication makes use of data products from the Two Micron All Sky Survey, which is a joint project of the University of Massachusetts and the Infrared Processing and Analysis Center/California Institute of Technology, funded by the National Aeronautics and Space Administration and the National Science Foundation.
This research has made use of the SIMBAD database and the VizieR catalogue access tool, CDS, Strasbourg, France
This research has made use of the NASA/IPAC Extragalactic Database (NED) which is operated by the Jet Propulsion Laboratory, California Institute of Technology, under contract with the National Aeronautics and Space Administration.

\end{acknowledgements}

%% file: chapter/table1.tex
\longtabL{1}{
\begin{landscape}
\begin{longtable}{|l|l|r|r|r|r|r|r|}
\caption{\label{tbl:ttsdata}AKARI, UCAC, and 2MASS photometric data for the previously known Taurus members.}\\

\hline
AKARI IRC name$^{\rm a}$ & Identifier & AKARI \textit{S9W}$^{\rm b}$ & AKARI \textit{L18W}$^{\rm b}$ & UCAC$^{\rm c}$ & 2MASS $J$ & 2MASS $H$ & 2MASS $K_S$\\
 & & [mag] & [mag] & [mag] & [mag] & [mag] & [mag]\\
\hline
\endfirsthead
\caption{continued.}\\
\hline
AKARI IRC name$^{\rm a}$ & Identifier & AKARI \textit{S9W}$^{\rm b}$ & AKARI \textit{L18W}$^{\rm b}$ & UCAC$^{\rm c}$ & 2MASS $J$ & 2MASS $H$ & 2MASS $K_S$\\
 & & [mag] & [mag] & [mag] & [mag] & [mag] & [mag]\\
\hline
\endhead
\hline
\endfoot

\input{chapter/table1_data}

\end{longtable}
$^{\rm a}$
Source name from its J2000 coordinates, following the IAU Recommendations for Nomenclature (2006).
The format is HHMMSSS$\pm$DDMMSS.
The source must be referred to in the literatures by its full name: AKARI-IRC-V1 J0123456+765432, where V1 refers to the version code.

$^{\rm b}$
The zero-magnitude flux densities for the AKARI bands are 56.262 Jy and 12.001 Jy for the \textit{S9W} and \textit{L18W} bands, respectively.
The AKARI measurements are based on the V-1 version of the IRC-PSC.

$^{\rm c}$
The UCAC magnitude is derived from aperture photometry.

$^{\rm d}$
1RXS J032409.7+123745 seems to be a giant star.
\end{landscape}
}

%% file: chapter/table1_data.tex
0311138+222457 & 1RXS J031113.0+222518     &  6.847 $\pm$ 0.154 &                    &  8.359 $\pm$ 0.093 &  7.273 $\pm$ 0.020 &  7.069 $\pm$ 0.053 &  6.968 $\pm$ 0.027\\
0324065+234706 & 1RXS J032406.6+234714     &  6.008 $\pm$ 0.048 &                    & 10.018 $\pm$ 1.000 &  7.128 $\pm$ 0.020 &  6.492 $\pm$ 0.020 &  6.267 $\pm$ 0.018\\
0324102+123746 & 1RXS J032409.7+123745$^{\rm d}$     &  2.950 $\pm$ 0.017 &  2.902 $\pm$ 0.023 &  6.841 $\pm$ 0.484 &  4.097 $\pm$ 0.292 &  3.490 $\pm$ 0.248 &  3.394 $\pm$ 0.244\\
0327325+255400 & 1RXS J032733.2+255405     &  6.947 $\pm$ 0.508 &                    &  9.090 $\pm$ 1.000 &  7.927 $\pm$ 0.032 &  7.550 $\pm$ 0.051 &  7.488 $\pm$ 0.023\\
0400095+081818 & RXJ0400.1+0818N           &  7.041 $\pm$ 0.183 &                    &  9.775 $\pm$ 0.048 &  8.281 $\pm$ 0.024 &  8.028 $\pm$ 0.000 &  7.843 $\pm$ 0.000\\
0404430+261856 & L1489 IRS                 &  3.745 $\pm$ 0.020 &  0.694 $\pm$ 0.022 &                    & 14.129 $\pm$ 0.000 & 11.908 $\pm$ 0.000 &  9.859 $\pm$ 0.043\\
0406513+254127 & RXJ0406.8+2541            &  6.207 $\pm$ 0.126 &                    & 11.452 $\pm$ 0.058 &  8.767 $\pm$ 0.021 &  8.026 $\pm$ 0.034 &  7.750 $\pm$ 0.018\\
0413271+281624 & Anon 1                    &  6.785 $\pm$ 0.221 &                    & 13.261 $\pm$ 0.171 &  8.826 $\pm$ 0.030 &  7.794 $\pm$ 0.027 &  7.460 $\pm$ 0.029\\
0413532+281124 & IRAS 04108+2803 A         &  6.022 $\pm$ 0.116 &                    &                    & 13.639 $\pm$ 0.024 & 11.521 $\pm$ 0.024 & 10.370 $\pm$ 0.019\\
0413573+291820 & IRAS 04108+2910           &  5.648 $\pm$ 0.044 &  3.559 $\pm$ 0.152 & 16.015 $\pm$ 0.316 & 11.310 $\pm$ 0.023 & 10.157 $\pm$ 0.024 &  9.356 $\pm$ 0.024\\
0414129+281212 & V773 Tau ABC              &  4.213 $\pm$ 0.028 &  2.247 $\pm$ 0.039 & 10.434 $\pm$ 0.033 &  7.494 $\pm$ 0.024 &  6.644 $\pm$ 0.020 &  6.210 $\pm$ 0.047\\
0414135+281249 & FM Tau                    &  5.729 $\pm$ 0.057 &  3.554 $\pm$ 0.036 &                    & 10.329 $\pm$ 0.021 &  9.391 $\pm$ 0.022 &  8.763 $\pm$ 0.022\\
0414146+282758 & FN Tau                    &  5.383 $\pm$ 0.096 &  2.780 $\pm$ 0.039 & 14.151 $\pm$ 0.267 &  9.469 $\pm$ 0.022 &  8.669 $\pm$ 0.036 &  8.189 $\pm$ 0.018\\
0414170+281058 & CW Tau                    &  4.144 $\pm$ 0.043 &  2.324 $\pm$ 0.049 & 13.390 $\pm$ 0.169 &  9.557 $\pm$ 0.022 &  8.243 $\pm$ 0.027 &  7.127 $\pm$ 0.023\\
0414176+280609 & CIDA-1                    &  6.126 $\pm$ 0.126 &  4.187 $\pm$ 0.156 &                    & 11.726 $\pm$ 0.021 & 10.584 $\pm$ 0.022 &  9.877 $\pm$ 0.021\\
0414263+280602 & IRAS 04113+2758 A         &  3.525 $\pm$ 0.017 &  1.031 $\pm$ 0.067 &                    & 12.475 $\pm$ 0.029 &  9.878 $\pm$ 0.033 &  7.777 $\pm$ 0.031\\
0414305+280514 & MHO 3                     &  4.049 $\pm$ 0.028 &  1.185 $\pm$ 0.022 &                    & 11.178 $\pm$ 0.021 &  9.249 $\pm$ 0.022 &  8.243 $\pm$ 0.023\\
0414478+264810 & CX Tau                    &  6.227 $\pm$ 0.093 &  3.725 $\pm$ 0.100 & 13.589 $\pm$ 0.073 &  9.867 $\pm$ 0.022 &  9.054 $\pm$ 0.024 &  8.807 $\pm$ 0.019\\
0414479+275234 & LkCa 3 AB                 &  6.860 $\pm$ 0.189 &                    & 11.865 $\pm$ 0.063 &  8.363 $\pm$ 0.032 &  7.625 $\pm$ 0.023 &  7.423 $\pm$ 0.021\\
0414492+281230 & FO Tau AB                 &  5.719 $\pm$ 0.105 &  3.511 $\pm$ 0.052 & 15.099 $\pm$ 0.192 &  9.650 $\pm$ 0.021 &  8.568 $\pm$ 0.021 &  8.121 $\pm$ 0.031\\
0417337+282047 & CY Tau                    &  6.270 $\pm$ 0.081 &                    & 13.022 $\pm$ 0.069 &  9.828 $\pm$ 0.020 &  8.970 $\pm$ 0.024 &  8.597 $\pm$ 0.021\\
0417496+282936 & V410 X-ray 1              &  5.997 $\pm$ 0.086 &  4.415 $\pm$ 0.049 & 16.680 $\pm$ 0.515 & 11.020 $\pm$ 0.021 &  9.733 $\pm$ 0.027 &  9.081 $\pm$ 0.019\\
0418107+251957 & [GBM90] L1506 1           &  6.537 $\pm$ 0.096 &                    & 13.039 $\pm$ 0.223 & 10.744 $\pm$ 0.024 &  9.589 $\pm$ 0.022 &  9.032 $\pm$ 0.019\\
0418310+282716 & V410 Tau ABC              &  7.184 $\pm$ 0.111 &                    & 10.366 $\pm$ 0.003 &  8.449 $\pm$ 0.018 &  7.789 $\pm$ 0.021 &  7.629 $\pm$ 0.026\\
0418311+281629 & DD Tau AB                 &  4.271 $\pm$ 0.027 &  2.366 $\pm$ 0.043 & 14.748 $\pm$ 0.232 &  9.830 $\pm$ 0.021 &  8.682 $\pm$ 0.020 &  7.878 $\pm$ 0.020\\
0418316+281658 & CZ Tau AB                 &  4.471 $\pm$ 0.033 &  2.461 $\pm$ 0.046 & 15.239 $\pm$ 0.023 & 10.517 $\pm$ 0.021 &  9.774 $\pm$ 0.022 &  9.362 $\pm$ 0.017\\
0418320+283115 & IRAS 04154+2823           &  4.882 $\pm$ 0.102 &  2.420 $\pm$ 0.048 &                    & 15.187 $\pm$ 0.048 & 12.369 $\pm$ 0.022 & 10.274 $\pm$ 0.017\\
0418344+283029 & V410 X-ray 2              &                    &  4.559 $\pm$ 0.071 &                    & 13.773 $\pm$ 0.026 & 10.708 $\pm$ 0.022 &  9.215 $\pm$ 0.017\\
0418406+281914 & V892 Tau                  &  1.278 $\pm$ 0.016 & -1.777 $\pm$ 0.016 & 14.846 $\pm$ 0.111 &  8.742 $\pm$ 0.027 &  7.016 $\pm$ 0.029 &  5.787 $\pm$ 0.016\\
0418470+282008 & Hubble 4                  &  6.931 $\pm$ 0.275 &                    & 12.181 $\pm$ 0.093 &  8.560 $\pm$ 0.020 &  7.636 $\pm$ 0.026 &  7.293 $\pm$ 0.024\\
0418514+282026 & CoKu Tau/1                &  5.239 $\pm$ 0.089 &  1.884 $\pm$ 0.022 &                    & 12.866 $\pm$ 0.030 & 11.489 $\pm$ 0.031 & 10.974 $\pm$ 0.024\\
0419010+281942 & V410 X-ray 6              &                    &  4.747 $\pm$ 0.066 &                    & 10.527 $\pm$ 0.022 &  9.597 $\pm$ 0.024 &  9.129 $\pm$ 0.021\\
0419128+282933 & FQ Tau AB                 &  7.175 $\pm$ 0.294 &                    & 15.097 $\pm$ 0.195 & 10.491 $\pm$ 0.026 &  9.702 $\pm$ 0.024 &  9.313 $\pm$ 0.022\\
0419158+290626 & BP Tau                    &  5.335 $\pm$ 0.123 &  3.303 $\pm$ 0.049 & 11.927 $\pm$ 0.081 &  9.098 $\pm$ 0.037 &  8.220 $\pm$ 0.024 &  7.736 $\pm$ 0.023\\
0419354+282721 & FR Tau                    &  6.853 $\pm$ 0.357 &                    & 16.321 $\pm$ 0.900 & 10.954 $\pm$ 0.021 & 10.374 $\pm$ 0.022 &  9.971 $\pm$ 0.018\\
0420258+281922 & SST Tau 042025.8+281923 &  6.564 $\pm$ 0.132 &  4.612 $\pm$ 0.054 &                    & 15.327 $\pm$ 0.039 & 13.422 $\pm$ 0.026 & 11.718 $\pm$ 0.021\\
0420260+280409 & SST Tau 042026.0+280408 &  6.643 $\pm$ 0.242 &  3.740 $\pm$ 0.071 & 14.696 $\pm$ 0.190 & 10.612 $\pm$ 0.021 &  9.949 $\pm$ 0.022 &  9.697 $\pm$ 0.018\\
0421079+270220 & CFHT-19                   &  5.309 $\pm$ 0.030 &  2.735 $\pm$ 0.054 &                    & 13.855 $\pm$ 0.024 & 12.062 $\pm$ 0.026 & 10.543 $\pm$ 0.021\\
0421432+193413 & IRAS 04187+1927           &  4.274 $\pm$ 0.041 &  2.494 $\pm$ 0.027 &                    & 10.185 $\pm$ 0.022 &  8.725 $\pm$ 0.024 &  8.021 $\pm$ 0.018\\
0421556+275506 & DE Tau                    &  5.382 $\pm$ 0.052 &  3.455 $\pm$ 0.017 & 12.773 $\pm$ 0.165 &  9.180 $\pm$ 0.022 &  8.273 $\pm$ 0.018 &  7.799 $\pm$ 0.018\\
0421574+282635 & RY Tau                    &  1.652 $\pm$ 0.017 & -0.273 $\pm$ 0.019 & 10.064 $\pm$ 0.037 &  7.155 $\pm$ 0.019 &  6.128 $\pm$ 0.061 &  5.395 $\pm$ 0.023\\
0421588+281806 & HD 283572                 &  6.805 $\pm$ 0.274 &                    &  8.847 $\pm$ 0.044 &  7.414 $\pm$ 0.029 &  7.008 $\pm$ 0.026 &  6.869 $\pm$ 0.023\\
0421594+193205 & T Tau N (+Sab)            &  1.539 $\pm$ 0.019 & -1.127 $\pm$ 0.029 &  9.747 $\pm$ 0.071 &  7.240 $\pm$ 0.023 &  6.237 $\pm$ 0.017 &  5.325 $\pm$ 0.017\\
0422007+265733 & FS Tau B                  &  5.327 $\pm$ 0.057 &  2.181 $\pm$ 0.316 &                    & 15.082 $\pm$ 0.082 & 13.351 $\pm$ 0.053 & 11.753 $\pm$ 0.029\\
0422022+265729 & FS Tau Aab                &  4.453 $\pm$ 0.044 &  1.877 $\pm$ 0.149 & 13.796 $\pm$ 0.900 & 10.705 $\pm$ 0.027 &  9.244 $\pm$ 0.026 &  8.178 $\pm$ 0.017\\
0422167+265456 & CFHT-21                   &  5.699 $\pm$ 0.065 &  3.793 $\pm$ 0.076 & 13.575 $\pm$ 1.000 & 11.577 $\pm$ 0.019 & 10.037 $\pm$ 0.022 &  9.011 $\pm$ 0.021\\
0422478+264552 & IRAS 04196+2638           &  5.759 $\pm$ 0.099 &  4.330 $\pm$ 0.053 &                    & 11.589 $\pm$ 0.022 & 10.147 $\pm$ 0.025 &  9.287 $\pm$ 0.019\\
0423077+280557 & IRAS 04200+2759           &                    &  3.998 $\pm$ 0.140 &                    & 13.179 $\pm$ 0.021 & 11.600 $\pm$ 0.024 & 10.413 $\pm$ 0.018\\
0423354+250300 & FU Tau                    &  6.199 $\pm$ 0.024 &                    &                    & 10.781 $\pm$ 0.026 &  9.945 $\pm$ 0.027 &  9.324 $\pm$ 0.024\\
0423391+245613 & FT Tau                    &  6.030 $\pm$ 0.071 &  3.809 $\pm$ 0.128 & 14.791 $\pm$ 0.174 & 10.192 $\pm$ 0.026 &  9.123 $\pm$ 0.027 &  8.596 $\pm$ 0.021\\
0424570+271156 & IP Tau                    &  6.019 $\pm$ 0.039 &                    & 12.046 $\pm$ 0.900 &  9.781 $\pm$ 0.021 &  8.893 $\pm$ 0.017 &  8.349 $\pm$ 0.018\\
0426535+260654 & FV Tau AB                 &  4.086 $\pm$ 0.027 &  2.116 $\pm$ 0.066 & 14.135 $\pm$ 0.900 &  9.917 $\pm$ 0.022 &  8.325 $\pm$ 0.024 &  7.442 $\pm$ 0.020\\
0426573+260629 & KPNO-13                   &  7.081 $\pm$ 0.157 &                    &                    & 11.281 $\pm$ 0.022 & 10.172 $\pm$ 0.022 &  9.580 $\pm$ 0.021\\
0427025+260531 & DG Tau B                  &  5.336 $\pm$ 0.027 &  1.838 $\pm$ 0.051 &                    & 16.516 $\pm$ 0.141 & 13.525 $\pm$ 0.000 & 12.038 $\pm$ 0.000\\
0427028+254222 & DF Tau AB                 &  4.272 $\pm$ 0.054 &  2.644 $\pm$ 0.074 & 11.703 $\pm$ 0.075 &  8.171 $\pm$ 0.026 &  7.256 $\pm$ 0.023 &  6.734 $\pm$ 0.024\\
0427046+260616 & DG Tau A                  &  2.995 $\pm$ 0.017 &  0.369 $\pm$ 0.026 &                    &  8.691 $\pm$ 0.018 &  7.722 $\pm$ 0.031 &  6.992 $\pm$ 0.020\\
0427572+261918 & IRAS 04248+2612 AB        &  6.750 $\pm$ 0.105 &  3.307 $\pm$ 0.045 &                    & 13.235 $\pm$ 0.053 & 11.795 $\pm$ 0.047 & 11.026 $\pm$ 0.040\\
0429208+274207 & IRAS 04262+2735           &  4.379 $\pm$ 0.036 &  3.326 $\pm$ 0.065 & 15.095 $\pm$ 1.000 &  8.550 $\pm$ 0.026 &  7.461 $\pm$ 0.016 &  7.015 $\pm$ 0.017\\
0429217+270126 & IRAS 04263+2654           &  6.545 $\pm$ 0.149 &  4.190 $\pm$ 0.107 &                    & 10.801 $\pm$ 0.022 &  9.496 $\pm$ 0.022 &  8.725 $\pm$ 0.017\\
0429237+243301 & GV Tau AB                 &  1.579 $\pm$ 0.023 & -1.163 $\pm$ 0.016 &                    & 11.544 $\pm$ 0.025 &  9.587 $\pm$ 0.023 &  8.054 $\pm$ 0.024\\
0429300+243955 & IRAS 04264+2433           &  5.838 $\pm$ 0.065 &  2.047 $\pm$ 0.056 &                    & 13.728 $\pm$ 0.036 & 11.982 $\pm$ 0.031 & 11.135 $\pm$ 0.027\\
0429360+243554 & J04293606+2435556         &  6.573 $\pm$ 0.041 &                    &                    & 10.780 $\pm$ 0.019 &  9.392 $\pm$ 0.020 &  8.659 $\pm$ 0.020\\
0429415+263258 & DH Tau AB                 &  6.332 $\pm$ 0.201 &  4.073 $\pm$ 0.070 & 13.462 $\pm$ 1.000 &  9.767 $\pm$ 0.021 &  8.824 $\pm$ 0.026 &  8.178 $\pm$ 0.026\\
0429515+260644 & IQ Tau                    &  5.274 $\pm$ 0.068 &  3.574 $\pm$ 0.070 & 13.890 $\pm$ 0.226 &  9.415 $\pm$ 0.020 &  8.417 $\pm$ 0.023 &  7.779 $\pm$ 0.023\\
0430039+181348 & UX Tau C                  &                    &  3.478 $\pm$ 0.038 &                    &  8.623 $\pm$ 0.023 &  7.960 $\pm$ 0.018 &  7.551 $\pm$ 0.021\\
0430296+242644 & FX Tau AB                 &  5.505 $\pm$ 0.079 &  3.556 $\pm$ 0.125 & 12.230 $\pm$ 0.900 &  9.388 $\pm$ 0.024 &  8.398 $\pm$ 0.018 &  7.924 $\pm$ 0.016\\
0430442+260124 & DK Tau AB                 &  4.191 $\pm$ 0.030 &  2.399 $\pm$ 0.033 & 11.986 $\pm$ 0.071 &  8.719 $\pm$ 0.030 &  7.758 $\pm$ 0.024 &  7.096 $\pm$ 0.016\\
0430502+230008 & IRAS 04278+2253           &  2.103 $\pm$ 0.019 &  0.198 $\pm$ 0.020 & 15.580 $\pm$ 0.364 &  8.778 $\pm$ 0.024 &  7.040 $\pm$ 0.015 &  5.855 $\pm$ 0.023\\
0430513+244221 & ZZ Tau AB                 &  6.831 $\pm$ 0.153 &                    & 14.544 $\pm$ 1.000 &  9.495 $\pm$ 0.021 &  8.695 $\pm$ 0.029 &  8.441 $\pm$ 0.021\\
0430517+244147 & ZZ Tau IRS                &  5.883 $\pm$ 0.069 &  2.903 $\pm$ 0.103 &                    & 12.842 $\pm$ 0.023 & 11.435 $\pm$ 0.026 & 10.314 $\pm$ 0.023\\
0431341+180804 & L1551 IRS5                &  2.794 $\pm$ 0.026 &                    &                    & 13.708 $\pm$ 0.060 & 11.505 $\pm$ 0.052 &  9.822 $\pm$ 0.035\\
0431361+181344 & LkHa 358                  &  5.437 $\pm$ 0.170 &                    &                    & 12.792 $\pm$ 0.032 & 10.922 $\pm$ 0.029 &  9.687 $\pm$ 0.021\\
0431384+181357 & HL Tau                    &  2.697 $\pm$ 0.030 & -0.231 $\pm$ 0.058 &                    & 10.624 $\pm$ 0.042 &  9.171 $\pm$ 0.046 &  7.410 $\pm$ 0.017\\
0431400+181357 & XZ Tau AB                 &  3.421 $\pm$ 0.079 &                    & 13.678 $\pm$ 0.124 &  9.385 $\pm$ 0.027 &  8.148 $\pm$ 0.040 &  7.291 $\pm$ 0.024\\
0431505+242418 & HK Tau AB                 &  6.255 $\pm$ 0.019 &  3.428 $\pm$ 0.060 & 13.588 $\pm$ 0.900 & 10.451 $\pm$ 0.022 &  9.253 $\pm$ 0.022 &  8.593 $\pm$ 0.018\\
0431578+182137 & V710 Tau AB               &  5.999 $\pm$ 0.282 &  4.013 $\pm$ 0.194 & 13.560 $\pm$ 1.000 &  9.279 $\pm$ 0.032 &  9.107 $\pm$ 0.030 &  8.654 $\pm$ 0.025\\
0432154+242859 & Haro 6-13                 &  4.501 $\pm$ 0.093 &  1.604 $\pm$ 0.061 &                    & 11.237 $\pm$ 0.024 &  9.319 $\pm$ 0.020 &  8.101 $\pm$ 0.029\\
0432303+173139 & GG Tau Aa                 &  4.540 $\pm$ 0.111 &  2.599 $\pm$ 0.051 & 11.785 $\pm$ 0.019 &  8.674 $\pm$ 0.035 &  7.815 $\pm$ 0.026 &  7.364 $\pm$ 0.018\\
0432317+242002 & FZ Tau                    &  4.138 $\pm$ 0.119 &  2.417 $\pm$ 0.030 & 13.352 $\pm$ 0.900 &  9.895 $\pm$ 0.022 &  8.400 $\pm$ 0.029 &  7.347 $\pm$ 0.017\\
0432320+225726 & IRAS 04295+2251           &                    &  1.947 $\pm$ 0.068 &                    & 14.889 $\pm$ 0.044 & 11.982 $\pm$ 0.036 & 10.141 $\pm$ 0.024\\
0432429+255231 & UZ Tau Aab                &  4.085 $\pm$ 0.022 &  2.247 $\pm$ 0.022 & 10.985 $\pm$ 1.000 &  9.136 $\pm$ 0.000 &  8.117 $\pm$ 0.000 &  7.354 $\pm$ 0.033\\
0432491+225303 & JH112                     &  5.189 $\pm$ 0.019 &  3.086 $\pm$ 0.047 & 15.093 $\pm$ 0.172 & 10.238 $\pm$ 0.027 &  8.995 $\pm$ 0.032 &  8.169 $\pm$ 0.023\\
0433062+240933 & GH Tau AB                 &  5.483 $\pm$ 0.137 &  3.804 $\pm$ 0.159 & 12.584 $\pm$ 1.000 &  9.109 $\pm$ 0.021 &  8.234 $\pm$ 0.027 &  7.794 $\pm$ 0.021\\
0433066+240955 & V807 Tau AB               &  5.198 $\pm$ 0.089 &  3.638 $\pm$ 0.073 & 11.197 $\pm$ 1.000 &  8.146 $\pm$ 0.023 &  7.357 $\pm$ 0.026 &  6.960 $\pm$ 0.016\\
0433190+224633 & IRAS 04303+2240           &  3.695 $\pm$ 0.060 &  2.000 $\pm$ 0.035 & 17.446 $\pm$ 1.000 & 11.103 $\pm$ 0.020 &  9.209 $\pm$ 0.023 &  7.673 $\pm$ 0.027\\
0433329+180059 & HD 28867                  &  5.092 $\pm$ 0.096 &  3.885 $\pm$ 0.256 &  6.466 $\pm$ 0.093 &  5.915 $\pm$ 0.035 &  5.813 $\pm$ 0.029 &  5.786 $\pm$ 0.023\\
0433341+242114 & GI Tau                    &  3.998 $\pm$ 0.127 &  1.848 $\pm$ 0.043 & 13.488 $\pm$ 1.000 &  9.341 $\pm$ 0.020 &  8.418 $\pm$ 0.021 &  7.888 $\pm$ 0.023\\
0433346+242106 & GK Tau                    &  4.478 $\pm$ 0.089 &                    & 12.044 $\pm$ 1.000 &  9.053 $\pm$ 0.027 &  8.108 $\pm$ 0.026 &  7.468 $\pm$ 0.021\\
0433367+260949 & IS Tau AB                 &  5.662 $\pm$ 0.056 &  3.970 $\pm$ 0.047 &                    & 10.323 $\pm$ 0.021 &  9.293 $\pm$ 0.023 &  8.642 $\pm$ 0.018\\
0433390+252038 & DL Tau                    &  4.749 $\pm$ 0.091 &  2.673 $\pm$ 0.041 & 12.894 $\pm$ 0.112 &  9.630 $\pm$ 0.021 &  8.679 $\pm$ 0.027 &  7.960 $\pm$ 0.021\\
0433394+175152 & HN Tau AB                 &  4.538 $\pm$ 0.061 &  2.574 $\pm$ 0.046 & 13.290 $\pm$ 0.124 & 10.699 $\pm$ 0.026 &  9.471 $\pm$ 0.027 &  8.384 $\pm$ 0.021\\
0433446+261500 & SST Tau 043344.6+261500 &  6.791 $\pm$ 0.116 &                    &                    & 11.639 $\pm$ 0.021 & 10.385 $\pm$ 0.022 &  9.744 $\pm$ 0.018\\
0433519+225030 & CI Tau                    &  5.034 $\pm$ 0.093 &  2.926 $\pm$ 0.024 & 13.157 $\pm$ 0.142 &  9.480 $\pm$ 0.020 &  8.431 $\pm$ 0.040 &  7.793 $\pm$ 0.020\\
0433546+261326 & IT Tau AB                 &  5.661 $\pm$ 0.091 &  3.911 $\pm$ 0.169 & 13.073 $\pm$ 0.900 &  9.866 $\pm$ 0.025 &  8.591 $\pm$ 0.036 &  7.860 $\pm$ 0.026\\
0434554+242852 & AA Tau                    &  5.439 $\pm$ 0.017 &  3.556 $\pm$ 0.025 & 12.373 $\pm$ 1.000 &  9.433 $\pm$ 0.024 &  8.546 $\pm$ 0.023 &  8.047 $\pm$ 0.024\\
0435273+241458 & DN Tau                    &  5.816 $\pm$ 0.063 &  3.647 $\pm$ 0.120 & 12.171 $\pm$ 0.087 &  9.139 $\pm$ 0.021 &  8.342 $\pm$ 0.027 &  8.015 $\pm$ 0.021\\
0435410+241108 & CoKu Tau/3 AB             &  5.129 $\pm$ 0.072 &  3.664 $\pm$ 0.043 &                    & 10.731 $\pm$ 0.026 &  9.197 $\pm$ 0.026 &  8.411 $\pm$ 0.024\\
0435473+225021 & HQ Tau                    &  4.208 $\pm$ 0.064 &                    & 11.788 $\pm$ 0.144 &  8.655 $\pm$ 0.024 &  7.731 $\pm$ 0.016 &  7.135 $\pm$ 0.021\\
0435528+225422 & HP Tau AB                 &  4.383 $\pm$ 0.090 &  2.049 $\pm$ 0.031 & 14.090 $\pm$ 0.201 &  9.549 $\pm$ 0.022 &  8.469 $\pm$ 0.065 &  7.625 $\pm$ 0.024\\
0435568+225436 & Haro 6-28 AB              &  6.885 $\pm$ 0.072 &                    & 15.986 $\pm$ 1.000 & 11.142 $\pm$ 0.022 & 10.055 $\pm$ 0.022 &  9.531 $\pm$ 0.020\\
0437514+262358 & HT Tau                    &  6.423 $\pm$ 0.143 &                    & 12.859 $\pm$ 1.000 &  8.635 $\pm$ 0.026 &  7.542 $\pm$ 0.018 &  7.129 $\pm$ 0.018\\
0438285+261048 & DO Tau                    &  3.965 $\pm$ 0.050 &  1.706 $\pm$ 0.062 &                    &  9.470 $\pm$ 0.022 &  8.243 $\pm$ 0.033 &  7.303 $\pm$ 0.017\\
0438352+261038 & HV Tau AB+C               &  7.162 $\pm$ 0.264 &                    &                    &  9.227 $\pm$ 0.023 &  8.284 $\pm$ 0.026 &  7.906 $\pm$ 0.024\\
0439174+224753 & VY Tau A                  &  6.376 $\pm$ 0.123 &                    & 13.385 $\pm$ 0.081 &  9.970 $\pm$ 0.023 &  9.260 $\pm$ 0.021 &  8.958 $\pm$ 0.020\\
0439178+222103 & LkCa 15                   &  5.834 $\pm$ 0.109 &  4.117 $\pm$ 0.123 & 11.875 $\pm$ 0.128 &  9.424 $\pm$ 0.020 &  8.600 $\pm$ 0.018 &  8.163 $\pm$ 0.018\\
0439208+254501 & GN Tau B                  &  5.405 $\pm$ 0.149 &  3.515 $\pm$ 0.027 & 15.465 $\pm$ 0.900 & 10.196 $\pm$ 0.025 &  8.893 $\pm$ 0.026 &  8.060 $\pm$ 0.027\\
0439557+254501 & IC2087 IRS                &  2.631 $\pm$ 0.077 &  0.839 $\pm$ 0.084 &                    & 10.668 $\pm$ 0.023 &  8.052 $\pm$ 0.018 &  6.275 $\pm$ 0.018\\
0440080+260524 & IRAS 04370+2559           &  5.278 $\pm$ 0.237 &  3.081 $\pm$ 0.361 &                    & 12.406 $\pm$ 0.023 & 10.248 $\pm$ 0.029 &  8.869 $\pm$ 0.018\\
0441168+283959 & CoKu Tau/4                &                    &  3.532 $\pm$ 0.069 & 14.771 $\pm$ 0.135 & 10.163 $\pm$ 0.030 &  9.077 $\pm$ 0.024 &  8.656 $\pm$ 0.019\\
0441387+255624 & IRAS 04385+2550           &  5.488 $\pm$ 0.046 &  2.745 $\pm$ 0.039 &                    & 11.849 $\pm$ 0.023 & 10.123 $\pm$ 0.022 &  9.200 $\pm$ 0.018\\
0442077+252310 & V955 Tau Ab               &  4.937 $\pm$ 0.041 &  3.270 $\pm$ 0.018 &                    &  9.811 $\pm$ 0.022 &  8.601 $\pm$ 0.021 &  7.942 $\pm$ 0.016\\
0442210+252033 & CIDA-7                    &  7.088 $\pm$ 0.058 &                    &                    & 11.397 $\pm$ 0.023 & 10.575 $\pm$ 0.022 & 10.169 $\pm$ 0.018\\
0442376+251537 & DP Tau                    &  4.898 $\pm$ 0.058 &  2.641 $\pm$ 0.056 & 15.058 $\pm$ 0.900 & 10.995 $\pm$ 0.021 &  9.689 $\pm$ 0.017 &  8.760 $\pm$ 0.016\\
0446530+165959 & DQ Tau                    &  5.093 $\pm$ 0.031 &  2.820 $\pm$ 0.065 & 13.253 $\pm$ 0.180 &  9.511 $\pm$ 0.021 &  8.544 $\pm$ 0.020 &  7.981 $\pm$ 0.021\\
0446590+170238 & Haro 6-37 A               &  4.597 $\pm$ 0.019 &  2.747 $\pm$ 0.064 & 13.655 $\pm$ 1.000 &  9.239 $\pm$ 0.028 &  7.991 $\pm$ 0.021 &  7.310 $\pm$ 0.024\\
0447062+165842 & DR Tau                    &  3.552 $\pm$ 0.020 &                    & 12.187 $\pm$ 0.410 &  8.845 $\pm$ 0.024 &  7.799 $\pm$ 0.053 &  6.874 $\pm$ 0.017\\
0447485+292511 & DS Tau                    &  5.621 $\pm$ 0.085 &  4.245 $\pm$ 0.050 & 12.490 $\pm$ 0.165 &  9.465 $\pm$ 0.018 &  8.597 $\pm$ 0.033 &  8.036 $\pm$ 0.029\\
0451473+304712 & UY Aur A                  &  3.479 $\pm$ 0.018 &  1.152 $\pm$ 0.016 & 12.151 $\pm$ 0.137 &  9.134 $\pm$ 0.020 &  7.987 $\pm$ 0.016 &  7.239 $\pm$ 0.018\\
0452066+304717 & IRAS 04489+3042           &  5.826 $\pm$ 0.067 &  3.246 $\pm$ 0.041 &                    & 14.426 $\pm$ 0.030 & 12.021 $\pm$ 0.021 & 10.383 $\pm$ 0.018\\
0452096+303744 & Haro 6-39                 &  7.023 $\pm$ 0.181 &  4.539 $\pm$ 0.110 &                    & 13.254 $\pm$ 0.021 & 12.117 $\pm$ 0.018 & 11.187 $\pm$ 0.018\\
0455095+182629 & HD 31281                  &  7.221 $\pm$ 0.257 &                    &  9.105 $\pm$ 0.088 &  7.974 $\pm$ 0.027 &  7.681 $\pm$ 0.017 &  7.609 $\pm$ 0.029\\
0455110+302159 & GM Aur                    &  6.793 $\pm$ 0.063 &  3.338 $\pm$ 0.131 &                    &  9.341 $\pm$ 0.018 &  8.603 $\pm$ 0.024 &  8.283 $\pm$ 0.017\\
0455458+303303 & AB Aur                    &  1.185 $\pm$ 0.016 & -0.841 $\pm$ 0.026 &  7.543 $\pm$ 0.157 &  5.936 $\pm$ 0.018 &  5.062 $\pm$ 0.020 &  4.230 $\pm$ 0.016\\
0455560+303622 & XEST26-062                &  6.521 $\pm$ 0.154 &  4.230 $\pm$ 0.089 & 15.339 $\pm$ 0.054 & 10.471 $\pm$ 0.021 &  9.660 $\pm$ 0.018 &  9.267 $\pm$ 0.019\\
0455593+303401 & SU Aur                    &  3.445 $\pm$ 0.016 &  0.671 $\pm$ 0.017 &  9.057 $\pm$ 0.027 &  7.199 $\pm$ 0.020 &  6.558 $\pm$ 0.020 &  5.990 $\pm$ 0.023\\
0457065+314250 & RXJ0457.0+3142            &  6.108 $\pm$ 0.055 &                    & 10.002 $\pm$ 0.081 &  7.486 $\pm$ 0.021 &  6.756 $\pm$ 0.018 &  6.538 $\pm$ 0.023\\
0503066+252319 & V836 Tau                  &  6.378 $\pm$ 0.297 &                    & 13.817 $\pm$ 0.203 &  9.913 $\pm$ 0.023 &  9.077 $\pm$ 0.029 &  8.595 $\pm$ 0.019\\
0505228+253131 & CIDA-9                    &  6.337 $\pm$ 0.062 &  4.870 $\pm$ 0.021 & 16.870 $\pm$ 1.000 & 12.808 $\pm$ 0.035 & 11.913 $\pm$ 0.043 & 11.161 $\pm$ 0.029\\
0507495+302404 & RW Aur A                  &  3.453 $\pm$ 0.047 &  1.620 $\pm$ 0.036 &  9.878 $\pm$ 0.076 &  8.378 $\pm$ 0.024 &  7.621 $\pm$ 0.038 &  7.020 $\pm$ 0.018\\
0529083+115212 & 1RXS J052908.4+115207     &  2.605 $\pm$ 0.017 &  0.168 $\pm$ 0.024 &  9.443 $\pm$ 0.049 &  7.698 $\pm$ 0.030 &  7.103 $\pm$ 0.029 &  6.590 $\pm$ 0.029\\
0529406+291110 & 1RXS J052940.9+291058     &  5.039 $\pm$ 0.037 &                    &  6.260 $\pm$ 0.265 &  5.314 $\pm$ 0.020 &  5.208 $\pm$ 0.252 &  5.028 $\pm$ 0.021\\
0536516+232605 & 1RXS J053652.7+232600     &  6.789 $\pm$ 0.209 &                    &  8.591 $\pm$ 0.032 &  7.328 $\pm$ 0.029 &  7.022 $\pm$ 0.021 &  6.913 $\pm$ 0.020\\
0537184+133452 & 1RXS J053718.4+133453     &  7.056 $\pm$ 0.016 &                    &  9.377 $\pm$ 0.030 &  8.105 $\pm$ 0.024 &  7.720 $\pm$ 0.020 &  7.593 $\pm$ 0.020\\

%% file: chapter/table2.tex
\longtabL{2}{
\begin{landscape}
\begin{longtable}{|l|r|r|r|r|r|r|}
\caption{\label{tbl:newdata}AKARI, UCAC, and 2MASS photometric data for the new TTS candidates.}\\

\hline
AKARI IRC name$^{\rm a}$ & AKARI \textit{S9W}$^{\rm b}$ & AKARI \textit{L18W}$^{\rm b}$ & UCAC$^{\rm c}$ & 2MASS $J$ & 2MASS $H$ & 2MASS $K_S$\\
 & [mag] & [mag] & [mag] & [mag] & [mag] & [mag]\\
\hline
\endfirsthead
\caption{continued.}\\
\hline
AKARI IRC name$^{\rm a}$ & AKARI \textit{S9W}$^{\rm b}$ & AKARI \textit{L18W}$^{\rm b}$ & UCAC$^{\rm c}$ & 2MASS $J$ & 2MASS $H$ & 2MASS $K_S$\\
 & [mag] & [mag] & [mag] & [mag] & [mag] & [mag]\\
\hline
\endhead
\hline
\endfoot

\input{chapter/table2_data}

\end{longtable}
$^{\rm a}$
Source name from its J2000 coordinates, following the IAU Recommendations for Nomenclature (2006).
The format is HHMMSSS$\pm$DDMMSS.
The source must be referred to in the literatures by its full name: AKARI-IRC-V1 J0123456+765432, where V1 refers to the version code.

$^{\rm b}$
The zero-magnitude flux densities for the AKARI bands are 56.262 Jy and 12.001 Jy for the \textit{S9W} and \textit{L18W} bands, respectively.
The AKARI measurements are based on the V-1 version of the IRC-PSC.

$^{\rm c}$
The UCAC magnitude is derived from aperture photometry.
\end{landscape}
}

%% file: chapter/table2_data.tex
0322025+305129 &  5.761 $\pm$ 0.077 &                    & 10.027 $\pm$ 0.022 &  8.227 $\pm$ 0.018 &  7.850 $\pm$ 0.020 &  7.663 $\pm$ 0.020\\
0325067+310652 &  6.473 $\pm$ 0.111 &  3.546 $\pm$ 0.125 & 12.613 $\pm$ 0.060 &  9.684 $\pm$ 0.022 &  8.930 $\pm$ 0.021 &  8.438 $\pm$ 0.018\\
0325125+305922 &  6.589 $\pm$ 0.154 &                    & 13.913 $\pm$ 0.086 & 10.243 $\pm$ 0.021 &  9.414 $\pm$ 0.021 &  9.067 $\pm$ 0.020\\
0337114+330303 &  6.441 $\pm$ 0.109 &  4.488 $\pm$ 0.108 & 13.630 $\pm$ 0.212 & 10.631 $\pm$ 0.023 &  9.746 $\pm$ 0.021 &  9.223 $\pm$ 0.018\\
0349290+345800 &  6.227 $\pm$ 0.097 &  4.400 $\pm$ 0.242 & 13.742 $\pm$ 0.114 & 10.933 $\pm$ 0.022 & 10.082 $\pm$ 0.016 &  9.504 $\pm$ 0.018\\
0413573+291820 &  5.648 $\pm$ 0.044 &  3.559 $\pm$ 0.152 & 16.015 $\pm$ 0.316 & 11.310 $\pm$ 0.023 & 10.157 $\pm$ 0.024 &  9.356 $\pm$ 0.024\\
0414187+115812 &  6.330 $\pm$ 0.180 &  4.291 $\pm$ 0.194 & 12.456 $\pm$ 0.104 & 10.128 $\pm$ 0.022 &  9.554 $\pm$ 0.022 &  9.205 $\pm$ 0.018\\
0421156+100722 &  6.732 $\pm$ 0.101 &                    & 11.685 $\pm$ 0.064 &  9.107 $\pm$ 0.020 &  8.409 $\pm$ 0.047 &  8.168 $\pm$ 0.026\\
0425412+353718 &                    &  3.786 $\pm$ 0.254 & 13.669 $\pm$ 0.112 & 11.172 $\pm$ 0.021 & 10.545 $\pm$ 0.022 & 10.069 $\pm$ 0.017\\
0430375+355031 &  6.264 $\pm$ 0.178 &  3.827 $\pm$ 0.241 & 16.213 $\pm$ 0.372 & 10.027 $\pm$ 0.021 &  8.874 $\pm$ 0.018 &  8.052 $\pm$ 0.019\\
0435254+341901 &  6.208 $\pm$ 0.063 &  4.073 $\pm$ 0.163 & 15.805 $\pm$ 0.339 & 11.566 $\pm$ 0.022 & 10.339 $\pm$ 0.028 &  9.503 $\pm$ 0.021\\
0450190+092328 &  4.337 $\pm$ 0.033 &  2.876 $\pm$ 0.032 & 12.323 $\pm$ 0.900 &  7.118 $\pm$ 0.024 &  6.253 $\pm$ 0.024 &  5.970 $\pm$ 0.024\\
0502405+192237 &  4.908 $\pm$ 0.029 &  2.959 $\pm$ 0.086 & 13.449 $\pm$ 0.224 &  8.119 $\pm$ 0.030 &  7.013 $\pm$ 0.017 &  6.576 $\pm$ 0.017\\
0511021+295926 &  6.684 $\pm$ 0.138 &  4.367 $\pm$ 0.045 & 13.091 $\pm$ 0.057 & 10.063 $\pm$ 0.035 &  9.320 $\pm$ 0.041 &  9.022 $\pm$ 0.032\\
0512342+255847 &  6.145 $\pm$ 0.206 &                    & 13.313 $\pm$ 0.129 & 10.462 $\pm$ 0.022 &  9.706 $\pm$ 0.022 &  9.411 $\pm$ 0.018\\
0516039+061852 &  6.807 $\pm$ 0.080 &                    & 14.728 $\pm$ 0.112 & 11.286 $\pm$ 0.022 & 10.463 $\pm$ 0.027 &  9.978 $\pm$ 0.018\\
0517259+070022 &  5.130 $\pm$ 0.045 &  3.156 $\pm$ 0.036 & 11.000 $\pm$ 0.026 &  9.469 $\pm$ 0.039 &  8.941 $\pm$ 0.055 &  8.424 $\pm$ 0.034\\
0519413+053842 &  5.107 $\pm$ 0.031 &  2.086 $\pm$ 0.023 &  8.977 $\pm$ 0.018 &  8.041 $\pm$ 0.023 &  7.706 $\pm$ 0.023 &  7.482 $\pm$ 0.024\\
0522456+225444 &                    &  2.467 $\pm$ 0.063 & 13.459 $\pm$ 0.114 &  9.641 $\pm$ 0.022 &  8.693 $\pm$ 0.026 &  8.391 $\pm$ 0.017\\
0525113+191547 &  6.331 $\pm$ 0.101 &                    & 14.599 $\pm$ 0.143 & 11.309 $\pm$ 0.021 & 10.416 $\pm$ 0.026 &  9.918 $\pm$ 0.018\\
0525519+345228 &  3.100 $\pm$ 0.077 &  1.520 $\pm$ 0.122 & 14.503 $\pm$ 0.155 &  9.621 $\pm$ 0.022 &  8.499 $\pm$ 0.020 &  7.745 $\pm$ 0.023\\
0527433+031309 &  7.023 $\pm$ 0.067 &  4.304 $\pm$ 0.066 & 12.163 $\pm$ 0.096 & 10.378 $\pm$ 0.023 &  9.849 $\pm$ 0.024 &  9.708 $\pm$ 0.025\\
0527493+064638 &  3.568 $\pm$ 0.119 &                    & 14.185 $\pm$ 0.231 &  8.312 $\pm$ 0.026 &  7.290 $\pm$ 0.057 &  6.462 $\pm$ 0.021\\
0529091+235902 &  5.960 $\pm$ 0.116 &  4.267 $\pm$ 0.107 & 14.452 $\pm$ 0.120 & 11.259 $\pm$ 0.024 & 10.368 $\pm$ 0.020 &  9.784 $\pm$ 0.018\\
0529592+121947 &                    &  4.078 $\pm$ 0.188 & 12.949 $\pm$ 0.119 & 10.593 $\pm$ 0.020 &  9.836 $\pm$ 0.021 &  9.391 $\pm$ 0.018\\
0536009+113339 &  6.718 $\pm$ 0.072 &                    & 11.543 $\pm$ 0.091 &  9.863 $\pm$ 0.026 &  9.263 $\pm$ 0.022 &  8.980 $\pm$ 0.023\\
0537531+372456 &  5.511 $\pm$ 0.070 &  3.949 $\pm$ 0.035 & 15.577 $\pm$ 0.228 & 10.543 $\pm$ 0.022 &  9.827 $\pm$ 0.026 &  9.313 $\pm$ 0.024\\
0539023+085612 &  6.947 $\pm$ 0.039 &                    & 11.170 $\pm$ 0.037 &  9.744 $\pm$ 0.024 &  9.157 $\pm$ 0.024 &  8.736 $\pm$ 0.021\\